\documentclass[preprint,proceedings]{rmaa}

\usepackage{paralist}

\usepackage{psfrag,color}

\SetYear{2002}
\SetConfTitle{Galaxy Evolution: Theory and Observations}

\title{A photometric method to determine supermassive black hole masses}

\author{Alister~W.~Graham\altaffilmark{1,2}, 
        Peter~Erwin\altaffilmark{2}, 
        Nicola~Caon\altaffilmark{2} 
        and Ignacio~Trujillo\altaffilmark{2}}

\altaffiltext{1}{Department of Astronomy, University of Florida, USA}
\altaffiltext{2}{Instituto de Astrof\'\i{}sica de Canarias, Spain.}

\shortauthor{Graham et al.}
\shorttitle{Black Hole Masses}

\fulladdresses{
\item P.\ Erwin, N.\ Caon and I.\ Trujillo: Instituto de Astrof\'\i{}sica de Canarias, La Laguna, E-38200 Tenerife, Spain.
\item Alister~W.~Graham: Department of Astronomy, University of Florida,
  PO Box 112055, Gainesville, FL 32611-2055, USA (Graham@astro.ufl.edu). 
}

\listofauthors{A. W. Graham, P. Erwin, N. Caon, I. Trujillo}
\indexauthor{Graham, A. W.}
\indexauthor{Erwin, P.}
\indexauthor{Caon, N.}
\indexauthor{Trujillo, I.}

\abstract{We report the discovery of a strong correlation between
the shape of a bulge's light-profile and the mass of its central
supermassive black hole ($M_{\rm bh}$).   We find that
$\log (M_{\rm bh}/M_{\sun}) = 2.91(\pm0.38)\log(n) +  6.37(\pm0.21)$,
where $n$ is the S\'ersic $r^{1/n}$ shape index.
This correlation is shown to be at least as strong as the relationship
between the logarithm of the stellar velocity dispersion and
$\log M_{\rm bh}$ and has comparable scatter.}

\resumen{Presentamos el descubrimiento de una estrecha correlaci\'on entre
la forma del perfil de brillo de los bulbos y la masa del agujero negro
supermasivo ($M_{\rm bh}$) que contienen en su centro. Encontramos que 
$\log (M_{\rm bh}/M_{\sun}) = 2.91(\pm0.38)\log(n) +  6.37(\pm0.21)$,
donde $n$ es el par\'ametro de forma de la ley de S\'ersic $r^{1/n}$.
Esta correlaci\'on es al menos tan fuerte como la relaci\'on encontrada
entre el logaritmo de la dispersi\'on de velocidades estelares y
$\log M_{\rm bh}$, y tiene una dispersi\'on comparable.}

\addkeyword{galaxies: fundamental parameters}
\addkeyword{galaxies: kinematics and dynamics}
\addkeyword{galaxies: nuclei}
\addkeyword{galaxies: structure}

\begin{document}
\maketitle

\begin{figure*}
  \includegraphics[width=\textwidth]{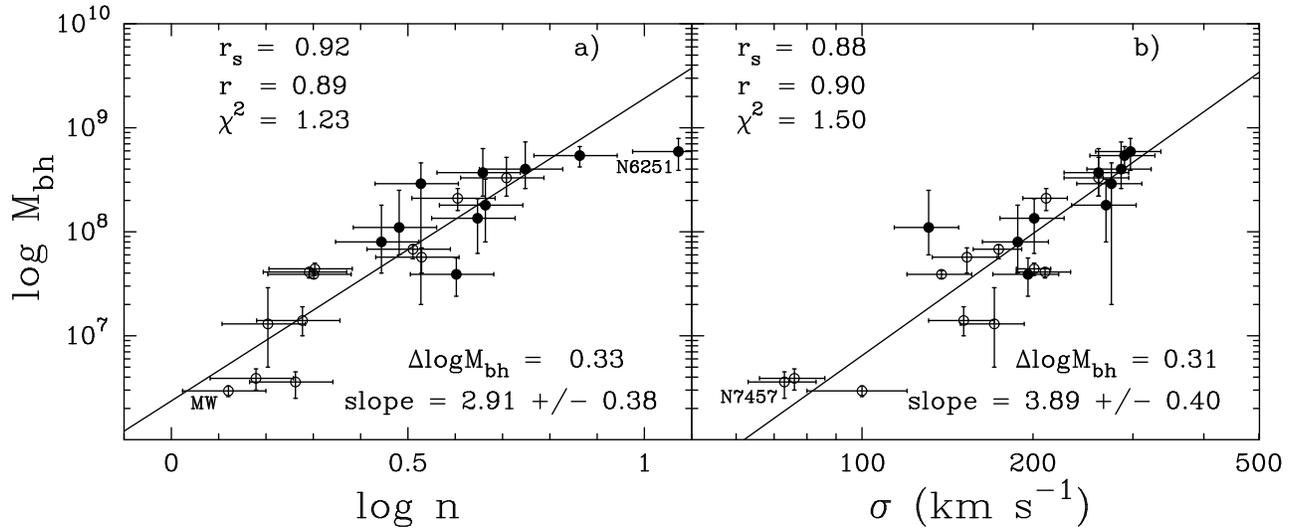}
  \caption{Correlations between the logarithm of a bulge's 
supermassive black hole mass and its 
a) bulge shape parameter (i.e.\ S\'ersic index $n$) 
and b) stellar velocity dispersion within $r_e/8$. 
The orthogonal linear-regression routine from Akritas \& Bershady (1996) 
has been used.  
The Spearman rank-order correlation coefficient $r_s$ is given, as is
the Pearson linear correlation coefficient $r$.
The $\chi^2$ merit function for a linear fit and the absolute
vertical scatter $\Delta \log M_{\rm bh}$ about the linear fit are also 
given.  
Elliptical galaxies are denoted by filled circles, lenticulars and
spirals by open circles.  
} 
  \label{fig:one}
\end{figure*}

\section{Data and results}

Our galaxy sample has come from the updated list of 
galaxies with supermassive black hole (SMBH) mass 
estimates given in the first two sections of Merritt 
\& Ferrarese's (2001) Table~1 
(see also Kormendy \& Gebhardt 2001 and Gebhardt et al.\ 2000).  
This initial sample of 30 galaxies was reduced to 22 because we
were unable to find reliable, and publicly available, images
for all 30 galaxies --- excluded galaxies were usually too 
large for a single CCD image to have sufficient sky background. 

The data, and reduction procedures, will be described 
in Erwin et al.\ (2002).  The extracted galaxy light profiles 
were modelled with either a seeing-convolved S\'ersic $r^{1/n}$
profile or, in cases when a disk was also present, with a 
seeing-convolved combination of S\'ersic bulge and exponential disk.  
The values of the bulge S\'ersic index $n$ 
were converted into the `central concentration index' $C_{r_e}(1/3)$ 
(Trujillo, Graham, \& Caon 2001; Graham, Trujillo, \& Caon 2001b) 
and plotted against the SMBH mass in Graham et al.\ (2001a).  
The current presentation, based on this work, shows the results 
from plotting $n$ directly against the SMBH mass (Figure 1a).

The orthogonal regression routine we have used in Figure 1  
treats both variables 
equally, and allows for intrinsic scatter as well as measurement 
errors in the data; as Merritt \& Ferrarese (2001) point out, 
it is generally the best method to use when there are errors in 
both variables.  
A 20\% error has been assumed for the value of $n$. The errors for the 
SMBH mass and central velocity dispersion $\sigma$ come from 
the above mentioned tables.  

A word of caution may be in order when comparing the different measures
of significance for the relations shown in Figure 1a and 1b.  
The strength of a correlation 
itself --- regardless of which function fits it --- is best measured
by the Spearman rank-order coefficient $r_{s}$.  The $\chi^{2}$ merit
function for a {\it linear fit} to the data, the Pearson coefficient $r$, 
and the vertical scatter in $\log M_{\rm bh}$ all measure how well a 
straight line fits the data (or the logarithm of the data, as the case 
may be).   

The $\chi^{2}$ value depends on the size of the measurement errors:
overestimating the errors will decrease the resulting $\chi^{2}$, even
though the correlation is unchanged; underestimating the errors can
produce a misleadingly large $\chi^{2}$.  Thus, even though the $\chi^{2}$
values for the $\log M_{\rm bh}$--$\log n$ relation are smaller than those for
the $\log M_{\rm bh}$--$\sigma_{c}$ relation, we do not take that as strong
evidence that the $\log M_{\rm bh}$--$\log n$ relation is better.
Ferrarese \& Merritt (2000) argued that their optimal
$\log M_{\rm bh}-\log \sigma_c$ relation had negligible intrinsic
scatter ($\chi^{2}<1$); this led them 
to posit that, ``Our results suggest that the stellar velocity dispersion
may be the fundamental parameter regulating the evolution of
supermassive BHs in galaxies.''   All twelve of their `Sample A'
galaxies, on which this conclusion was based, had an uncertainty
of $\pm13$\% on their central velocity dispersions, except for the
Milky Way which had an uncertainty of $\pm20$\%.
If these uncertainties have been overestimated, it will result in an
underestimate to the $\chi ^2$ value of the fit which may then lead
one to wrongly conclude that there is no intrinsic scatter in the
relation.  The situation is identical if the 20\% errors we assigned
to the S\'ersic indices are too large.  However, irrespective of the 
errors one assigns, the strengths of both correlations shown in Figure 1 
appear to be equal.  {\it This suggests that one can use (relatively 
inexpensive) photometric images, instead of velocity dispersion 
measurements, for determining SMBH masses.}

It is well established that more luminous bulges have larger values 
of $n$.  They also have greater central concentrations, deeper
gravitational potential wells and higher central gravitational 
potential gradients (Ciotti 1991, Trujillo et al.\ 2002).  
One might expect these characteristics to result in bulges 
more able to fuel and build their central black holes.
To date, most models incorporating SMBHs
have addressed their formation from either the standpoint of 
the older $\log M_{\rm bh}$--$M_{\rm bulge}$ relation 
or the more recent $\log M_{\rm bh}$--$\log \sigma$ 
correlation (Figure 1b). 
It is hoped 
\adjustfinalcols
that a more complete understanding will be achieved 
when the correlation between
SMBH mass and bulge light-profile shape is additionally explained.

\end{document}